\documentclass[conference,10pt]{IEEEtran}
\IEEEoverridecommandlockouts
\usepackage[utf8]{inputenc}
\usepackage[T1]{fontenc}
\usepackage{amsmath,amssymb,amsfonts}
\usepackage{bm}
\usepackage{graphicx}
\usepackage{xcolor}
\usepackage{tikz}
\usetikzlibrary{positioning,calc,arrows.meta,fit,backgrounds}
\usepackage{booktabs}
\usepackage{array}
\usepackage{ragged2e}
\usepackage{caption}
\usepackage{makecell}
\usepackage{hyperref}
\usepackage{cleveref}
\usepackage{balance}
\usepackage{enumitem}
\usepackage{tabularx}
\usepackage{fontawesome5}

\newcolumntype{L}[1]{%
  >{\RaggedRight\arraybackslash\hspace{0pt}}p{#1}%
}

\newcolumntype{C}[1]{%
  >{\Centering\arraybackslash}p{#1}%
}

\newcolumntype{I}{%
  >{\Centering\arraybackslash}m{0.62cm}%
}

\newcolumntype{Y}{%
  >{\RaggedRight\arraybackslash\hspace{0pt}}X%
}

\setlist{nosep,leftmargin=*}

\graphicspath{{figures/}}

\begin{document}

\title{Incentives and Evidence in Learned Service Orchestration}

\author{
  \IEEEauthorblockN{
    Syed Izhan Khilji\IEEEauthorrefmark{2},
    Alireza Furutanpey$^*$\thanks{\textsuperscript{*}Corresponding author}%
      \IEEEauthorrefmark{2}\IEEEauthorrefmark{3},
    Schahram Dustdar\IEEEauthorrefmark{2}
  }
  \IEEEauthorblockA{
    \IEEEauthorrefmark{2}Distributed Systems Group (DSG),
    TU Wien, Vienna, Austria\\
    \{i.khilji, dustdar\}@dsg.tuwien.ac.at
  }

  \IEEEauthorblockA{
    \IEEEauthorrefmark{3}Coovally.ai, Barcelona, Spain\\
    a.furutanpey@coovally.ai
  }
}
\maketitle

\begin{abstract}
Reinforcement learning for service orchestration has been the subject of sustained research for over a decade, yet it is not used in production at scale. The usual explanation is that learned controllers degrade under delayed and noisy telemetry, workload shifts, and uncontrolled tenants. We test whether existing evidence supports that explanation. We evaluate three highly influential RL-based orchestration systems spanning resource allocation, DAG scheduling, and autoscaling, using pre-registered predictions about comparative degradation under production-relevant perturbations and paired inference with family-wise error correction. Across the tests, most predicted performance reversals do not occur. Diagnostic analyses show that these outcomes often reflect comparator collapse, artefact limitations, or evaluation choices rather than evidence that learned controllers tolerate the perturbations. One apparent advantage under observation lag is roughly fortyfold compared to a Kubernetes HPA-equivalent controller. Another widely cited result cannot be reconstructed from its released artefact, and the strongest reproducible margin is far smaller than the published results. Conclusions also reverse under changes in perturbation magnitude and evaluation mode. Based on these results and broader patterns in the literature, we identify an institutional problem. Publication and review incentives favour benchmark gains against convenient comparators, even when those gains provide little evidence of deployment performance. We argue that the problem is not solely technical. Rather, it is institutional, so learned orchestration needs production-grade comparators, registered perturbation models, separate operational metrics, and publication criteria that reward reproducible operational evidence. Without these changes, the literature can grow without establishing whether learning improves orchestration.
\end{abstract}


\section{Introduction}
\label{sec:intro}

Service orchestration coordinates the runtime decisions that keep a distributed system running, such as how many replicas of a service to run, where to place them, how to route requests among them, and how to divide finite compute under shifting demand. These decisions set cost, latency, and reliability, and they are made continuously as the workload changes. A controller can make them reactively, from the state it observes, or it can anticipate where demand is going and act before it arrives. Whether the second, learned from data, beats the first has been an open question for over a decade.

Production control is reactive and hand-tuned. Google's Autopilot scales from moving-window statistics and a supervised recommender~\cite{rzadca2020autopilot}. Borg packs tasks by priority under engineered rules~\cite{verma2015large,burns2016borg}. The default scalers in Kubernetes, the most widely deployed orchestrator, compare one utilisation signal against a threshold. These controllers are robust and cheap to run, but they act on the present observation, optimise one step at a time, and are retuned by hand for each deployment.
Reinforcement learning is the dominant proposed alternative for a concrete reason. A hand-tuned rule does not adapt to the workload in front of it, acts one step at a time, and cannot express the coupled high-dimensional choices, which job on which machine, that scheduling and placement require. A learned policy can do all three. It anticipates rather than reacts, it is trained on the workload rather than tuned by hand, and it can represent decisions no simple rule captures. The published methods report large simulator improvements over hand-tuned heuristics for resource allocation~\cite{mao2016resource}, cluster scheduling~\cite{mao2019learning}, and auto-scaling~\cite{rossi2019horizontal}. That promise has sustained a decade of work.
Learned orchestration has not reached production with a noticeable impact. Production environments expose learned policies to noisy and delayed telemetry, biased reward proxies, abrupt workload shifts, and uncontrolled co-tenants. Each condition can degrade learned-policy performance~\cite{huang2017adversarial,zhang2020robust,rakhsha2020policy,behzadan2017vulnerability}. The unresolved question is whether these conditions disadvantage learned policies relative to the controllers operators actually deploy.

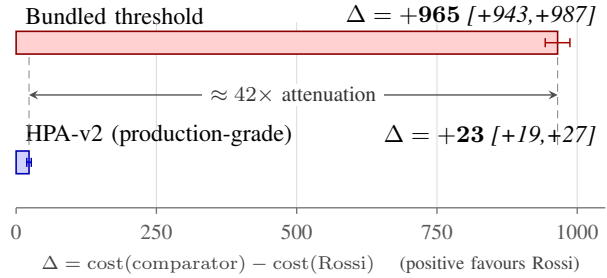
\begin{figure}[!t]
\centering
\begin{tikzpicture}[
  x=\columnwidth,
  y=1cm,
  axis/.style={draw=black!55, line width=0.5pt},
  grid/.style={draw=black!14, line width=0.35pt},
  guide/.style={draw=black!45, line width=0.4pt, densely dashed},
  tick/.style={font=\footnotesize, text=black!65, anchor=north},
]
\useasboundingbox (0,0) rectangle (1,4.15);

\def\xL{0.08}
\def\xR{0.96}
\pgfmathsetmacro{\xW}{\xR-\xL}
\def\scaleMax{1050}

\def\headOneY{3.58}
\def\barOneY{3.08}
\def\arrowY{2.54}
\def\headTwoY{2.00}
\def\barTwoY{1.50}
\def\axisY{1.05}
\def\barH{0.30}

\pgfmathsetmacro{\xTwentyThree}{\xL+\xW*23/\scaleMax}
\pgfmathsetmacro{\xNineSixtyFive}{\xL+\xW*965/\scaleMax}

\pgfmathsetmacro{\xNineteen}{\xL+\xW*19/\scaleMax}
\pgfmathsetmacro{\xTwentySeven}{\xL+\xW*27/\scaleMax}

\pgfmathsetmacro{\xNineFortyThree}{\xL+\xW*943/\scaleMax}
\pgfmathsetmacro{\xNineEightySeven}{\xL+\xW*987/\scaleMax}

\foreach \v in {250,500,750,1000}{%
  \pgfmathsetmacro{\xx}{\xL+\xW*\v/\scaleMax}
  \draw[grid]
    (\xx,\axisY)
    --
    (\xx,\barOneY+\barH+0.12);
}

\node[anchor=west,font=\small]
  at (\xL,\headOneY)
  {Bundled threshold};

\node[anchor=east,align=right,font=\small]
  at (\xR,\headOneY)
  {$\Delta=\mathbf{+965}$\;\textit{[+943,+987]}};

\fill[
  red!18,
  draw=red!55!black,
  line width=0.6pt
]
  (\xL,\barOneY)
  rectangle
  (\xNineSixtyFive,\barOneY+\barH);

\draw[red!65!black,line width=0.55pt]
  (\xNineFortyThree,\barOneY+0.15)
  --
  (\xNineEightySeven,\barOneY+0.15)

  (\xNineFortyThree,\barOneY+0.09)
  --
  (\xNineFortyThree,\barOneY+0.21)

  (\xNineEightySeven,\barOneY+0.09)
  --
  (\xNineEightySeven,\barOneY+0.21);

\node[anchor=west,font=\small]
  at (\xL,\headTwoY)
  {HPA-v2 (production-grade)};

\node[anchor=east,align=right,font=\small]
  at (\xR,\headTwoY)
  {$\Delta=\mathbf{+23}$\;\textit{[+19,+27]}};

\fill[
  blue!18,
  draw=blue!70!black,
  line width=0.6pt
]
  (\xL,\barTwoY)
  rectangle
  (\xTwentyThree,\barTwoY+\barH);

\draw[blue!75!black,line width=0.55pt]
  (\xNineteen,\barTwoY+0.15)
  --
  (\xTwentySeven,\barTwoY+0.15)

  (\xNineteen,\barTwoY+0.09)
  --
  (\xNineteen,\barTwoY+0.21)

  (\xTwentySeven,\barTwoY+0.09)
  --
  (\xTwentySeven,\barTwoY+0.21);

\draw[guide]
  (\xTwentyThree,\barOneY-0.04)
  --
  (\xTwentyThree,\barTwoY+\barH+0.04);

\draw[guide]
  (\xNineSixtyFive,\barOneY-0.04)
  --
  (\xNineSixtyFive,\barTwoY+\barH+0.04);

\draw[
  <->,
  >=stealth,
  draw=black!70,
  line width=0.5pt
]
  (\xTwentyThree,\arrowY)
  --
  node[
    midway,
    fill=white,
    inner xsep=3pt,
    inner ysep=1pt,
    font=\footnotesize,
    text=black!80
  ]
  {$\approx 42\times$ attenuation}
  (\xNineSixtyFive,\arrowY);

\draw[axis]
  (\xL,\axisY)
  --
  (\xR,\axisY);

\foreach \v in {0,250,500,750,1000}{%
  \pgfmathsetmacro{\xx}{\xL+\xW*\v/\scaleMax}

  \draw[axis]
    (\xx,\axisY)
    --
    (\xx,\axisY-0.08);

  \node[tick]
    at (\xx,\axisY-0.10)
    {\v};
}

\node[
  anchor=north,
  font=\scriptsize,
  text=black!75
]
  at (0.52,0.55)
  {$\Delta=
    \mathrm{cost}(\mathrm{comparator})
    -
    \mathrm{cost}(\mathrm{Rossi})$
    \quad (positive favours Rossi)};

\end{tikzpicture}
\caption{
Comparator choice attenuates Rossi's apparent advantage under $k=10,\mathrm{s}$ observation lag from $\Delta=+965$ against the bundled threshold controller to $+23$ against an HPA-v2-equivalent controller, approximately $42\times$. Positive $\Delta$ favours Rossi; whiskers show 95\% paired-bootstrap confidence intervals over 30 windows.
}

\label{fig:overview}
\end{figure}

Current evidence cannot determine whether learning improves orchestration for two reasons. First, deep reinforcement learning results are sensitive to implementation and evaluation choices~\cite{engstrom2020implementation,agarwal2021precipice}, and an advantage measured against a weak baseline may not hold up against a controller that operators would actually deploy. Second, existing evaluations rarely expose learned controllers to delayed telemetry, heavy-tailed workloads, or adversarial observation perturbations under reproducible protocols. Learned policies are vulnerable to adversarial inputs~\cite{huang2017adversarial,behzadan2017vulnerability,gleave2020adversarial}, while orchestration evaluations rarely test them under such inputs. They also omit the heavy-tailed, bursty, and skewed workloads observed in production traces~\cite{reiss2012heterogeneity,cortez2017resource} and addressed explicitly in large-scale system design~\cite{dean2013tail}.
%
We demonstrate the first problem by re-evaluating the seminal work by Rossi et al. for autoscaling~\cite{rossi2019horizontal} under a 10s observation lag using 30 paired workload windows. We hold the learned controller, workload windows, perturbation, and cost metric fixed and change only the comparator. The HPA-v2-equivalent controller implements a tolerance deadband and scale-down stabilisation; its implementation and configuration sensitivity are detailed in \Cref{sec:exp:hpa}. Against the unstabilised threshold controller bundled with the simulator, Rossi appears better by 965 cost units. Against the HPA-v2-equivalent controller, the estimated advantage falls to 23 units, an attenuation of approximately $42\times$ (\Cref{fig:overview}). The HPA-v2-equivalent controller is also cheaper in the clean, heavy-tailed, and adversarial conditions. The large lag advantage is therefore primarily evidence of comparator collapse rather than evidence that the learned controller degrades less under delayed telemetry than a production-representative autoscaler.

This result motivates our central question: \emph{do current evaluation practices distinguish lower learned-policy degradation under specified perturbations from comparator failure, artefact limitations, and evaluation choices}? We address this question by re-evaluating DeepRM~\cite{mao2016resource}, Decima~\cite{mao2019learning}, and Rossi~\cite{rossi2019horizontal}, which span resource allocation, DAG scheduling, and container autoscaling. We pre-register nine directional predictions that the learned policies will degrade more than their comparators under production-relevant perturbations and evaluate them using paired statistical inference with family-wise error correction (\Cref{sec:evidence}). Seven of the nine predictions fail, but these failures do not establish that the learned policies tolerate the tested perturbations. For DeepRM and Rossi, observation lag degrades the bundled comparators more than the learned policies, so the comparisons primarily measure comparator failure. Decima's released artefact cannot reconstruct the testbed comparison supporting its reported $21\%$ improvement, and the margin against the strongest comparator exposed by the artefact is $3.0\%$. The conclusions are also sensitive to evaluation choices: Decima's lag verdict reverses at one perturbation magnitude (\Cref{sec:exp:magnitude}), and Rossi's lag verdict reverses between frozen-checkpoint and online evaluation.

Our core contribution is to show why current evidence cannot determine whether learning improves orchestration and to identify the minimum requirements for a valid test. Learned policies must be compared with deployed controllers under production perturbations using reproducible protocols and operational outcomes. Until those requirements are met, reported gains do not establish progress.
\section{Background}
\label{sec:background}

We introduce the terminology and action-space perspective that the remainder of this work builds on.
\subsection{Reinforcement learning}
\label{sec:bg:rl}

A reinforcement-learning agent performs actions in an environment modelled as a Markov decision process $\mathcal{M}=(\mathcal{S},\mathcal{A},P,r,\gamma)$, with states $\mathcal{S}$, actions $\mathcal{A}$, transition kernel $P(s'\mid s,a)$, reward $r(s,a,s')$, and discount $\gamma\in(0,1)$~\cite{sutton2018reinforcement}. It learns a policy $\pi(a\mid s)$ that maximises the expected discounted return $\mathbb{E}[\sum_t \gamma^t r_t]$. Orchestration methods span value-based families such as, tabular~\cite{watkins1992qlearning} and deep (DQN~\cite{mnih2015humanlevel}, Double-DQN~\cite{hasselt2016deep}), policy gradients (PPO~\cite{schulman2017proximal}), and actor-critic methods (SAC~\cite{haarnoja2018soft}). These families, as standardly formulated, assume a stationary or slowly drifting process, a scalar extrinsic reward, and a tractable action space. Orchestration violates all three.

\subsection{Orchestration through its actions}
\label{sec:bg:actions}

The literature characterises an orchestration task by its state and reward. The action space is more informative. An orchestration action is a call to a control plane, with a precondition, a cost, and an effect that are fixed in advance. A replica cannot be removed when none exist. A pod cannot be placed on a node without capacity. Routing weights must sum to one. The action space therefore encodes what a policy can do and which formulations are well posed.

\begin{figure}[htb]
\centering
\begin{tikzpicture}[
  font=\footnotesize,
  >={Latex[length=1.6mm]},
  base/.style={rounded corners=2pt, draw, align=center, inner sep=4pt},
  dyn/.style={base, fill=black!7, draw=black!50, dashed, text width=1.85cm},
  act/.style={base, draw=black!60, text width=2.95cm, align=left},
  pol/.style={base, fill=black!82, text=white, text width=1.2cm, minimum height=0.95cm},
  lab/.style={align=center, font=\scriptsize\itshape, text=black!55}
]
\node[pol] (pi) {policy\\$\pi(a\!\mid\! s)$};
\node[dyn, left=0.65cm of pi] (dyn) {\textbf{dynamics}\\$P(s'\!\mid\! s,a)$};
\node[act, right=0.65cm of pi] (act) {\textbf{actions}\ $\mathcal{A}$\\[2pt]
  scale\ \ $\{-1,0,+1\}$\\
  place\ \ $\{0,1\}^{n\times m}$\\
  route\ \ $\Delta^{n-1}$};
\draw[->] (dyn) -- node[above,font=\scriptsize]{$s$} (pi);
\draw[->] (pi) -- node[above,font=\scriptsize]{$a$} (act);
\node[lab, below=2pt of dyn, text width=1.95cm] {opaque;\\learned from data};
\node[lab, below=2pt of act, text width=2.95cm] {typed by contract;\\known in advance};
\end{tikzpicture}
\caption{%
Orchestration provides asymmetric prior knowledge. System dynamics must be inferred from data, whereas the local semantics and feasibility constraints of actions are specified by the control-plane API. Generic RL formulations typically do not encode this action-space structure.%
}
\label{fig:actions}
\end{figure}
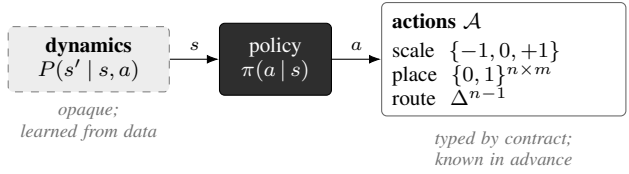

The two halves of the problem are not equally hard to model. The dynamics are opaque, because the effect of an action propagates through queueing, contention, and tenant behaviour that no simulator reproduces fully, so a model of them must be learned from data. The actions are not opaque, because their local semantics are given by API contract and are known before any data is collected (\Cref{fig:actions}). A standard formulation ignores this difference and learns both halves from data with no prior, which places the one structure the problem hands us, the action space, on the side that must be inferred. Domain knowledge enters most cheaply on the action side.

\subsection{Learning for systems}
\label{sec:bg:systems}

Reinforcement learning for cluster management and scheduling was advanced by Mao \emph{et al.}~\cite{mao2016resource,mao2019learning}, with antecedents in hybrid autonomic control~\cite{tesauro2006hybrid}, classification-driven cluster management~\cite{delimitrou2014quasar}, and a broad networking survey by Luong \emph{et al.}~\cite{luong2019applications}. The Park benchmark~\cite{mao2019park} collects twelve systems problems under one interface and notes how learning for systems differs from learning for games. The three methods we study are the most-cited of this line, but the practices we examine are not unique to them. More recent microservice systems, such as FIRM~\cite{qiu2020firm} and Sinan~\cite{zhang2021sinan}, compare against bespoke baselines in a simulator or a single cluster, the arrangement we examine here. Across the line, learning beats heuristics in simulation but has not displaced the production elasticity heuristics surveyed in~\cite{lorido2014review,qu2018autoscaling}, even where the deployment infrastructure exists.

\section{Orchestration Tasks and the Limits of Current Evidence}
\label{sec:taxonomy}

We first organise the task classes by the properties that affect whether reinforcement learning can improve control over heuristic or supervised baselines, namely action-space structure, observability, feedback delay, reversibility, and operational risk. We then argue that the field's evidence cannot establish whether it does.

\subsection{A problem-centred taxonomy}
\label{sec:taxonomy:classes}

\begin{table*}[htb]
\centering

\caption{%
Problem-centred taxonomy of RL for service orchestration by action-space structure, observability, feedback delay, reversibility, operational risk, and production status. The final column indicates whether RL has displaced heuristic or ML-augmented control in production.%
}
\label{tab:taxonomy}

\footnotesize
\setlength{\tabcolsep}{3pt}
\renewcommand{\arraystretch}{1.12}

\begin{tabularx}{\textwidth}{@{}
    >{\centering\arraybackslash}m{0.55cm}
    >{\RaggedRight\arraybackslash}p{2.20cm}
    >{\RaggedRight\arraybackslash}p{3.25cm}
    >{\RaggedRight\arraybackslash}p{2.00cm}
    >{\RaggedRight\arraybackslash}p{2.20cm}
    >{\RaggedRight\arraybackslash}p{1.85cm}
    >{\RaggedRight\arraybackslash}p{1.10cm}
    >{\RaggedRight\arraybackslash}X
@{}}

\toprule

&
\mbox{\textbf{Task class}}
&
\mbox{\textbf{Action-space structure}}
&
\mbox{\textbf{State signal}}
&
\mbox{\textbf{Observation lag}}
&
\mbox{\textbf{Reversibility}}
&
\mbox{\textbf{Risk}}
&
\mbox{\textbf{Production RL status}}
\\

\midrule

{\Large\faServer}
&
Horizontal/vertical\newline scaling
&
Discrete adjustment\newline
$\{-1,0,+1\}$
&
Utilisation,\newline queue depth
&
seconds--minutes
&
partial
&
medium
&
ML-augmented heuristic\newline
(Autopilot)
\\

\addlinespace[0.45em]

{\Large\faSitemap}
&
Load balancing\newline and routing
&
Continuous simplex\newline
$\Delta^{n-1}$
&
Rates,\newline response times
&
seconds
&
high
&
low
&
Heuristic control in production
\\

\addlinespace[0.45em]

{\Large\faChartPie}
&
Resource allocation
&
Constrained continuous\newline
$\mathbb{R}_{\geq 0}^{m\times d}$
&
Demands,\newline capacity
&
seconds--minutes
&
partial
&
high
&
Simulation evidence only
\\

\addlinespace[0.45em]

{\Large\faMapMarker}
&
Service placement\newline and migration
&
Binary assignment\newline
$\{0,1\}^{n\times m}$
&
Topology,\newline affinity
&
minutes--hours
&
low
&
high
&
Heuristic control in production
\\

\addlinespace[0.45em]

{\Large\faSlidersH}
&
SLO-driven control
&
Mixed discrete and continuous
controll knobs
&
SLO metrics
&
variable
&
very low
&
very high
&
Largely underexplored
\\

\bottomrule

\end{tabularx}
\end{table*}

\Cref{tab:taxonomy} profiles the task classes that account for most published RL for orchestration work. Each class is defined by the action-space structure, the state it observes, how long it waits before the effect of an action is visible, how reversible its actions are, and the operational risk of taking a wrong action.
The table points to three implications. First, difficulty rises on two axes at once. Standard algorithms are designed for complex decision problems or non-stationary environments, and handling both together remains hard~\cite{padakandla2021survey}, yet orchestration presents both. Second, the classes where reinforcement learning has its most credible results, load balancing and simple scaling, are the classes where conventional heuristics already perform well~\cite{lorido2014review,qu2018autoscaling}, so the benefit rarely justifies the operational cost of a learned component~\cite{rzadca2020autopilot,sculley2015hidden}. Third, as \Cref{sec:bg:actions} argued, the action space is more tractable than the dynamics, yet standard formulations treat the two alike and approximate every action with no structural prior. Each class has a distinct action geometry, monotone discrete, continuous simplex, or combinatorial assignment, that a generic parameterisation discards.

\subsection{Comparators are selected for tractability}
\label{sec:problem}
Across this literature, the baselines are chosen for tractability, not for the controllers operators run. The Rossi auto-scaler~\cite{rossi2019horizontal} ships with a simulator that bundles a threshold controller, the control class Kubernetes deploys by default, yet its published comparisons are against static deployments and other learned variants, and the bundled threshold is left unevaluated. Decima~\cite{mao2019learning} reports its $21\%$ improvement against tuned heuristics on a live Spark testbed, with Graphene among its strongest simulator baselines, yet its released simulator exposes static partitioning and FIFO as the classical comparators and ships no executable Graphene, so the artefact can reproduce neither comparison. The pattern is not particular to orchestration. In recommender systems, most published neural methods fail to beat classical baselines once those baselines are tuned~\cite{dacrema2019progress}, and within reinforcement learning itself, simple linear policies match deep methods on the benchmarks that established them~\cite{mania2018simple}. A field that selects comparators for availability inherits this failure mode by default. \Cref{sec:evidence} measures the cost. 
The Rossi advantage is attenuated by roughly fortyfold against a production-grade controller, and the Decima artefact cannot verify the claim for which it is cited.
\subsection{Production analogues of adversarial RL failure modes}
\label{sec:problem:conditions}
Each failure mode attributed to orchestration matches an adversarial machine learning attack class with a documented failure for deep reinforcement learning, and the preconditions of each hold in production with no adversary present. The modes are non-stationarity from workload drift and abrupt shifts in demand, the simulation-to-reality mismatch from simulators that omit contention and tail-latency dynamics~\cite{zhao2020simtoreal}, reward misspecification from biased proxies, exploration risk in irreversible settings~\cite{garcia2015comprehensive}, amplified by exploration bonuses that reward novelty itself~\cite{pathak2017curiosity,burda2019exploration}, and cluster heterogeneity that gives the same nominal task a different MDP on every cluster~\cite{reiss2012heterogeneity,cortez2017resource,dustdar2023continuum}. \Cref{tab:adversarial} states the correspondence.

\begin{table}[!t]
\centering
\caption{Each common orchestration phenomenon matches an adversarial machine learning attack class with a documented failure for deep reinforcement learning. The final column marks whether the corresponding prediction is tested in \Cref{sec:evidence}.}
\label{tab:adversarial}
\footnotesize
\setlength{\tabcolsep}{4pt}
\renewcommand{\arraystretch}{1.25}
\begin{tabular}{@{}p{0.35cm}p{2.3cm}p{2.5cm}p{1.5cm}@{}}
\toprule
\textbf{\#} & \textbf{Orchestration phenomenon} & \textbf{Adversarial-ML analogue} & \textbf{Tested} \\
\midrule
1   & Noisy / lagged telemetry          & Observation perturbation~\cite{huang2017adversarial,zhang2020robust} & Yes (P1) \\
1$^\star$ & Hostile observation noise   & Targeted adversarial input~\cite{huang2017adversarial}               & Yes (P3) \\
2   & SLOs as proxy rewards             & Reward poisoning~\cite{rakhsha2020policy}                            & No (\Cref{sec:discussion}) \\
3   & Co-tenant interference            & Adversarial co-agents~\cite{gleave2020adversarial}                   & No \\
4   & Abrupt workload shifts            & Environment poisoning~\cite{rakhsha2020policy,behzadan2017vulnerability}               & Yes (P2) \\
\bottomrule
\end{tabular}
\end{table}

Each row yields a directional prediction that, under realistic perturbation, the learned policy degrades more than the baseline. We test three, each instantiated on all three methods, giving the nine cells of \Cref{sec:evidence}.

\textbf{P1 (observation lag).} Production telemetry is sampled and aggregated with delay, and per-counter jitter is comparable to the perturbation budgets in adversarial RL~\cite{huang2017adversarial,zhang2020robust}. We predict that learned orchestrators degrade under realistic lag, and more than a non-learning baseline on the same telemetry.

\textbf{P2 (workload tail).} A release, a marketing event, or a tenant onboarding changes the transition kernel without changing the agent's representation of it, the operational form of environment poisoning~\cite{rakhsha2020policy,behzadan2017vulnerability} and of evasion through distribution shift~\cite{zhang2020robust}. We predict that a policy trained on one workload distribution degrades on another in proportion to the distance between them. We instantiate this as a service-time shift from exponential to Pareto-tailed ($\alpha=1.5$), calibrated to the Google cluster workload~\cite{reiss2012heterogeneity}, and for Decima as a reweighting of its DAG workload toward heavier-tailed production size statistics.

\textbf{P3 (adversarial observation).} A bounded FGSM-style perturbation of the policy input~\cite{goodfellow2014fgsm} should not only change the action but reverse the aggregate outcome.

Two rows are not tested. Co-tenant interference is the weakest analogue. Co-tenants are independent workloads competing for shared resources, not adversaries optimised against the orchestrator, so the correspondence is distribution shift rather than adversarial optimisation, and we name it as the next perturbation class to register. Reward poisoning points to something deeper than a robustness result. An SLO is an operational invariant the system must hold~\cite{nastic2020sloc}, and encoding it as one term of a scalar reward to be traded off against others is a category error, independent of any noise on the measurement~\cite{casamayor2025distributed}. We return to reward design in \Cref{sec:discussion}, and the Rossi cell in \Cref{sec:evidence} shows how a scalar weighting can make a comparison uninterpretable.

Two kinds of threat cut across these rows, and the field's evaluations apply neither. One is deliberate. An adversary chooses inputs to drive the policy to a costly action, and the P3 cell is the analogue, in which a bounded perturbation corrupts the policy's decisions but, in these simulators, not the aggregate outcome (\Cref{sec:exp:pattern}). The other is emergent. It arises from the workload itself with no adversary, and it is the more common. Load skew, heavy-tailed service times, hot keys, and bursty correlated arrivals are properties of real workloads that production traces report as the norm~\cite{reiss2012heterogeneity,cortez2017resource}, and that large-scale systems are explicitly engineered to tolerate~\cite{dean2013tail}. The robust-MDP framework formalises this as optimisation over an uncertainty set of such patterns~\cite{iyengar2005robust,nilim2005robust,wiesemann2013robust}, and the P2 cell is one point from that set, whose weak and method-specific outcome (\Cref{sec:exp:pattern}) marks the class as under-explored rather than safe. Orchestration benchmarks instead use benign synthetic arrivals, Poisson and exponential, so a method can look strong while never facing the worst-case structure it must survive.

The conditions for failure, naturally occurring, deliberately adversarial, and emergent, are present and almost entirely untested. Whether they produce the predicted degradation is the question of \Cref{sec:evidence}. The argument is not any single correspondence but the cumulative one, that production orchestration, by accident, assembles the conditions under which deep reinforcement learning is documented to fail~\cite{huang2017adversarial,zhang2020robust,behzadan2017vulnerability}.
\section{Empirical Evidence}
\label{sec:evidence}

We test the nine predictions of \Cref{sec:problem:conditions} on the three methods in \Cref{tab:methods}, each a canonical example of orchestration RL in one task class. Each is tested against the strongest non-learning comparator its released artefact ships, for two reasons. Where the artefact implements the published comparison, as for DeepRM, this tests the published claim on its own terms. Where it does not, as for Decima, it is the strongest comparison the released evidence permits, and \Cref{sec:exp:repro} documents the narrowing. A separate, pre-registered re-evaluation against a production-grade controller follows in \Cref{sec:exp:hpa}. The primary result is the difference observed between the two comparator classes, not the failure of the initial predictions, and this comparison was specified before any perturbed run rather than constructed after a null result.

\begin{table}[!t]
\centering
\caption{The three methods, and the simulator-grade comparator bundled with each. The limitations of these comparators as proxies for production controllers are the subject of \Cref{sec:exp:hpa}.}
\label{tab:methods}
\footnotesize
\setlength{\tabcolsep}{4pt}
\renewcommand{\arraystretch}{1.25}
\begin{tabular}{@{}p{1.6cm}p{2.0cm}p{3.6cm}@{}}
\toprule
\textbf{Method} & \textbf{Task class} & \textbf{Bundled comparator} \\
\midrule
DeepRM~\cite{mao2016resource} & Resource allocation & Tetris$^\star$, the packing heuristic of~\cite{grandl2014multi} as implemented in the DeepRM source, and SJF \\
Rossi~\cite{rossi2019horizontal} & Horizontal / vertical scaling & Threshold controller in the RLAD simulator, with no stabilisation features \\
Decima~\cite{mao2019learning} & DAG cluster scheduling & \texttt{dynamic\_partition}, static partitioning in the public simulator \\
\bottomrule
\end{tabular}
\end{table}

\paragraph{Protocol}
A prediction is \emph{pre-registered} if its (method, perturbation, anchor) cell, comparator, metric, perturbation magnitude, directional claim, statistical test, and rejection rule were written down and frozen in a timestamped pre-registration document before any perturbed run for that cell. For each cell we draw 30 paired evaluation seeds (DeepRM, Decima) or 30 non-overlapping windows of the official workload (Rossi), and compute $\Delta_i=\text{metric}_i(\text{comparator})-\text{metric}_i(\text{RL})$, where the metric is the one defined by the original paper (mean slowdown, total cost, mean job completion time). A positive $\Delta$ means the learned method is better, and the predicted degradation is $\Delta<0$. Confidence intervals are 95\% paired percentile bootstraps over 5000 resamples. The reported $p$-values are paired sign-flip values over $10^5$ flips, one-sided in the observed direction, corrected across the nine predictions by Holm--Bonferroni at $\alpha=0.05$, following the call for statistically defensible deep-RL evaluation~\cite{agarwal2021precipice}. Anchors are calibrated to the literature. P1 lag is set at $k=10\,$s from Borg telemetry P95 values~\cite{verma2015large}, and at $\lambda=1.0$ for Decima, the upper limit of the range for which its simulator is calibrated. P2 is set at Pareto $\alpha=1.5$, and for Decima at DAG-size tail weight $w=0.5$, which reweights its TPC-H workload toward the heavier-tailed DAG statistics of the Alibaba production trace. P3 is an FGSM perturbation at $\varepsilon=0.05$. P1 and P2 perturb the environment and apply to the learned policy and the comparator identically. P3 targets the learned policy's observation alone while the comparator reads the true state, the threat model in which the attacker corrupts the representation the defender computes on~\cite{huang2017adversarial}. The pre-registration document, perturbation specifications, controller implementations, and evaluation scripts accompany this submission as supplementary material and are released publicly\footnote{https://github.com/izhan19717/Learned-Service-Orchestration.git}.

\subsection{Reproduction gates}
\label{sec:exp:repro}

A perturbation test is informative only on a method whose clean behaviour reproduces. DeepRM passes a strict 30-seed gate with no silent admission drops, reaching mean slowdown $36.5$ against Tetris$^\star$ $61.0$ ($\Delta=+24.5$, CI $[+22.7,+26.3]$), with the authors' source-aligned values ($19.1$, $23.3$, $44.8$) also reproduced under the original protocol. Rossi matches all six metrics of its Table~I within sub-percent relative error, the worst case being $0.36\%$, well inside the pre-registered $15\%$ gate. The gate also fixes the evaluation mode. Rossi is a tabular controller that learns online, and a frozen-checkpoint variant fails the clean gate, posting cost $608.9$ against the reproduced $204.1$, and reverses the P1 verdict ($\Delta=-233.7$ at $k=10$, against $+965.1$ online). We therefore evaluate the published online controller and release the frozen sweep as a superseded sensitivity. A verdict that flips with the evaluation mode is itself the implementation-dependence documented for deep reinforcement learning~\cite{henderson2018rlmatters,engstrom2020implementation}, here surfacing in the comparison rather than the training. Decima is the exception. Trained under the authors' released code for the full $10{,}000$-epoch budget, our checkpoint improves on the README-exposed \texttt{dynamic\_partition} comparator by $3.0\%$ in mean JCT, against the $21\%$ the paper reports over tuned heuristics on a 25-node Spark testbed. The released artefact is the event-driven simulator, not the testbed. Its reference evaluation exposes static partitioning and FIFO as the classical schemes and ships no executable Graphene, so neither the headline comparison nor the strongest-baseline one can be reconstructed from it. We therefore report the Decima cells as a narrowed comparison, consistent with the broader finding that implementation choices and ablated comparators explain much of the difference between published RL results and independent re-implementations~\cite{engstrom2020implementation}.

\subsection{Outcomes}
\label{sec:exp:outcomes}

\Cref{tab:predictions} reports every cell. Seven of nine predictions are falsified. The two confirmations (Rossi P2, P3) are not perturbation-induced crossovers. The comparator already beats the learned controller in the clean cell ($\Delta_{\text{clean}}=-85.8$ on all 30 windows), and the perturbation only widens an existing divergence.

\begin{table*}[!t]
\centering
\caption{The nine pre-registered predictions and their anchor outcomes.
$\Delta=\text{metric}(\text{comparator})-\text{metric}(\text{RL})$,
with the per-method metric named in the second column. A positive
$\Delta$ means the learned method is better, and the predicted
degradation is $\Delta<0$. The quantity $p_{\mathrm{H}}$ is the
Holm--Bonferroni adjusted $p$-value across the family of nine.}
\label{tab:predictions}

\footnotesize
\setlength{\tabcolsep}{5pt}
\renewcommand{\arraystretch}{1.25}

\makebox[\textwidth][c]{%
\begin{tabular}{
    @{}
    l
    l
    >{\raggedright\arraybackslash}p{2.75cm}
    l
    r
    r
    r
    l
    @{}
}
\toprule
\textbf{ID}
& \textbf{Method (metric)}
& \textbf{Perturbation}
& \textbf{Anchor}
& \multicolumn{1}{c}{$\boldsymbol{\Delta}$}
& \multicolumn{1}{c}{\textbf{95\% CI}}
& \multicolumn{1}{c}{$\boldsymbol{p_{\mathrm{H}}}$}
& \textbf{Outcome}
\\
\midrule

P1-DR
& DeepRM (slowdown)
& Observation lag
& $k=10$
& $+195.7$
& $[+185.1,+205.9]$
& $<10^{-3}$
& falsified
\\

P2-DR
& DeepRM (slowdown)
& Workload tail
& $\alpha=1.5$
& $+91.8$
& $[+79.9,+104.6]$
& $<10^{-3}$
& falsified
\\

P3-DR
& DeepRM (slowdown)
& FGSM observation
& $\varepsilon=0.05$
& $+23.6$
& $[+21.7,+25.4]$
& $<10^{-3}$
& falsified
\\

\midrule

P1-Ro
& Rossi (cost)
& Observation lag
& $k=10\,\mathrm{s}$
& $+965.1$
& $[+943.1,+986.5]$
& $<10^{-3}$
& falsified$^\dagger$
\\

P2-Ro
& Rossi (cost)
& Service-time tail
& $\alpha=1.5$
& $-106.0$
& $[-113.1,-98.7]$
& $<10^{-3}$
& confirmed (widens)
\\

P3-Ro
& Rossi (cost)
& Bucket-flip
& $\varepsilon=0.05$
& $-91.0$
& $[-95.4,-86.8]$
& $<10^{-3}$
& confirmed (widens)
\\

\midrule

P1-De
& Decima (JCT)
& Observation lag
& $\lambda=1.0$
& $+716{,}396$
& $[+653{,}206,+779{,}842]$
& $<10^{-3}$
& falsified
\\

P2-De
& Decima (JCT)
& Workload tail
& $w=0.5$
& $+8{,}044$
& $[+1{,}138,+16{,}525]$
& $0.033$
& falsified$^{\ddagger}$
\\

P3-De
& Decima (JCT)
& FGSM node features
& $\varepsilon=0.05$
& $+2{,}001$
& $[+1{,}057,+3{,}019]$
& $<10^{-2}$
& falsified
\\

\bottomrule
\end{tabular}%
}

\vspace{2pt}

\parbox{\textwidth}{%
\footnotesize
$^\dagger$ The magnitude is a property of the bundled threshold.
Against an HPA-equivalent controller it drops by about $40\times$,
and the verdict survives only marginally
(\Cref{sec:exp:hpa}).
$^{\ddagger}$ Survives Holm at the boundary under the one-sided
convention, but not under the two-sided sensitivity convention
($p_{\mathrm{H,2s}}\approx 0.066$).
}

\end{table*}

\begin{figure}[!t]
\centering
\includegraphics[width=\columnwidth]{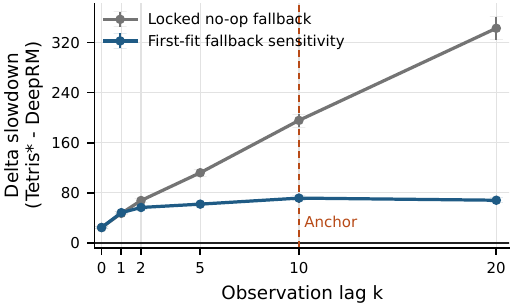}
\caption{The DeepRM P1 magnitude is largely an artefact of how a stale action is injected. Under the pre-registered no-op fallback (grey) the Tetris$^\star$ comparator falls through to a no-op and its cost grows without bound in the lag. Under a first-fit fallback (blue) the same comparator stays competent, and the difference at the $k=10$ anchor (dashed) falls by about $63\%$. The directional verdict is unchanged across the grid, but the magnitude is not. $\Delta=\mathrm{slowdown}(\mathrm{Tetris}^\star)-\mathrm{slowdown}(\mathrm{DeepRM})$.}
\label{fig:deeprm-p1-sens}
\end{figure}

In none of the three methods is the learned policy robust enough for an operator to rely on. For Rossi, the $+965$ P1 magnitude is the bundled threshold's bang-bang oscillation under stale telemetry, a baseline artefact that attenuates roughly $40\times$ against a stabilisation-windowed controller (\Cref{sec:exp:hpa}). For DeepRM, the magnitude is sensitive to the stale-action injection rule, under which the Tetris$^\star$ comparator falls through to a no-op in $68.6\%$ of perturbed steps. An alternative injection protocol reduces it by $63\%$ (\Cref{fig:deeprm-p1-sens}), so the cell reflects the perturbation harness as much as the method. For Decima, neither explanation applies. At the calibrated anchor, its lag-robustness survives even a stronger dependency-aware comparator (\Cref{sec:exp:hpa}), though \Cref{sec:exp:magnitude} shows that this particular verdict is sensitive to the lag magnitude, and the difficulty the Decima cells expose is not a weak baseline but a published comparison that cannot be reproduced from the released artefact. The remaining cells, dissected in \Cref{sec:exp:pattern}, divide along the same line, between artefacts of weak comparators and method behaviour that the released code does not let an independent party verify. Three falsified P1 predictions rest on three different causes, and none of them supports the conclusion that the method is sound. The count is uninformative because the comparisons behind it are.

\subsection{The cross-method pattern}
\label{sec:exp:pattern}

Three observations recur, each carrying the caveats above.

\emph{Observation lag breaks the weaker comparators harder than the learned policies, but not uniformly and not always because the comparator is weak.} P1 is falsified in all three methods. For Rossi and DeepRM the mechanism is the comparator. The Rossi threshold lacks production stabilisation, and the Tetris$^\star$ no-op fall-through under the injection rule inflates the DeepRM magnitude. Decima is the exception that disciplines the claim, because at the calibrated anchor its lag advantage holds even against a stronger dependency-aware scheduler (\Cref{sec:exp:hpa}), although that advantage is itself magnitude-sensitive (\Cref{sec:exp:magnitude}). We therefore do not claim that lag uniformly breaks heuristics, nor that learned policies are uniformly robust. What the three cells share is narrower. In no case does the bundled comparison let a reader conclude how the method would fare against a controller an operator would run.

\emph{Adversarial perturbations land on the decision, not the outcome.} P3 is falsified in all three, and the diagnostic is consistent. The attack bites at the policy output, changing DeepRM's argmax in $69.8\%$ of states in the locked P3 diagnostic ($75.2\%$ in the re-evaluation pass of the action-redundancy ablation, \Cref{sec:exp:ablation}), dropping Decima's target-action probability from $0.341$ to $0.279$, and flipping Rossi's bucketed control on $20.9\%$ of steps, yet it does not reverse the aggregate metric. The locus of the apparent robustness is therefore the action space rather than the state encoding, in every method. This locates \emph{where} the robustness sits, not yet \emph{why}. The ablation in \Cref{sec:exp:ablation} tests one candidate mechanism and does not support it, and the open question that remains motivates the action-space view of \Cref{sec:bg:actions}.

\begin{figure}[!t]
\centering
\includegraphics[width=\columnwidth]{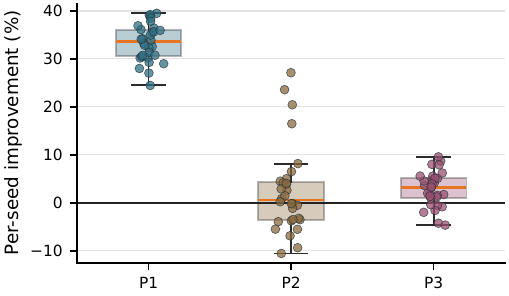}
\caption{Per-seed paired improvement for the three Decima cells, against \texttt{dynamic\_partition}. P1 lag is positive on all 30 seeds, ranging from $24\%$ to $40\%$. P2 tail has a median near zero with a few large positive outliers carrying the aggregate, the pattern behind its boundary rejection and the reason it generalises in neither direction. P3 is narrow and small. Positive means Decima is the cheaper policy.}
\label{fig:decima-dist}
\end{figure}

\emph{Workload-tail effects are weak and method-specific.} P2 is the closest to a mixed result. DeepRM rejects with a large positive $\Delta$. Rossi degrades as predicted, but only on top of pre-existing clean dominance. Decima rejects at the boundary, on an aggregate driven by a small number of outlier seeds, with only 16 of 30 seeds favouring Decima (\Cref{fig:decima-dist}). The class generalises in neither direction.

\subsection{Re-evaluation against stronger comparators}
\label{sec:exp:hpa}

The Rossi P1 result rests on a baseline whose bang-bang collapse is partly an artefact of the bundled implementation. To test whether it survives a comparator with production-grade stabilisation, we implemented a Kubernetes HPA-equivalent controller following the \texttt{autoscaling/v2} specification, with a proportional recommendation under a $10\%$ tolerance deadband, a $300\,$s scale-down stabilisation window, immediate scale-up, default rate policies, a $15\,$s sync period, replica bounds $[1,10]$, and a $50\%$ target utilisation. We re-ran the four Rossi cells under the same paired protocol. Three pre-implementation sanity gates confirmed the controller matches the specification before any perturbation cell was run.

\begin{table}[!t]
\centering
\caption{Re-evaluating the four Rossi cells against a production-grade HPA-equivalent controller. $\Delta=\text{cost}(\text{HPA-v2})-\text{cost}(\text{Rossi})$, where negative means HPA-v2 is cheaper. The bundled-threshold $\Delta$ is shown for comparison.}
\label{tab:hpa}
\footnotesize
\setlength{\tabcolsep}{4pt}
\renewcommand{\arraystretch}{1.25}
\begin{tabular}{@{}lrrl@{}}
\toprule
\textbf{Cell} & \multicolumn{1}{c}{$\bm{\Delta}$ \textbf{(bundled)}} & \multicolumn{1}{c}{$\bm{\Delta}$ \textbf{(HPA-v2)}} & \textbf{95\% CI (HPA-v2)} \\
\midrule
Clean                          & $-85.8$  & $-31.6$ & $[-36.4,-26.8]$ \\
P1 lag $k=10\,$s               & $+965.1$ & $+23.0$ & $[+18.8,+27.2]$ \\
P2 tail $\alpha=1.5$           & $-106.0$ & $-52.4$ & $[-60.0,-45.0]$ \\
P3 bucket-flip $\varepsilon=0.05$ & $-91.0$ & $-36.8$ & $[-41.7,-32.5]$ \\
\bottomrule
\end{tabular}
\end{table}

\begin{figure}[!t]
\centering
\includegraphics[width=\columnwidth]{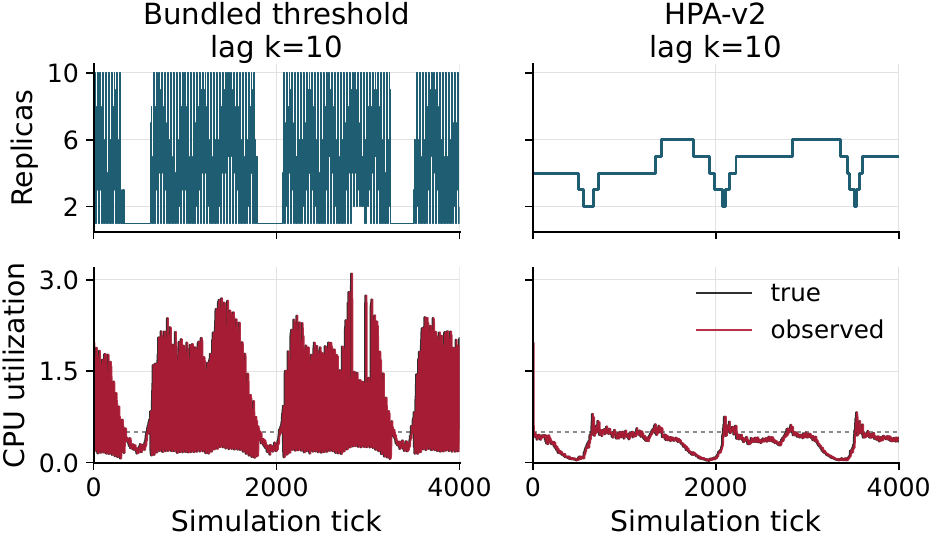}
\caption{The mechanism behind the $40\times$ difference, under observation lag $k=10\,$s. The bundled threshold (left) oscillates bang-bang between the replica bounds, with true utilisation repeatedly spiking to roughly $3\times$ overload. The HPA-equivalent controller (right) tracks the same workload within a narrow replica band and, after the initial transient, maintains a near-$50\%$ utilisation. 
The HPA-v2 configuration uses a $300$s scale-down stabilisation window and a $10\%$ tolerance deadband, which suppress scale-down responses to isolated lagged-low measurements.}
\label{fig:hpa-trace}
\end{figure}

\Cref{tab:hpa} reports the outcome. The lag-induced difference shrinks by roughly $40\times$ when the bundled threshold is replaced by HPA-v2, and \Cref{fig:hpa-trace} shows why. The production-grade controller does not oscillate under lag. The directional verdict survives in the narrow sense that Rossi is cheaper on all 30 lagged windows, but the magnitude is now the same order as its clean-cell advantage rather than an order larger. Across the four cells HPA-v2 wins three (clean, tail, adversarial) at $-31.6$ to $-52.4$ cost units, and loses only under lag. The published Rossi advantage therefore survives against an operationally realistic baseline under one perturbation class, and only by a small margin.

\begin{figure}[!t]
\centering
\includegraphics[width=\columnwidth]{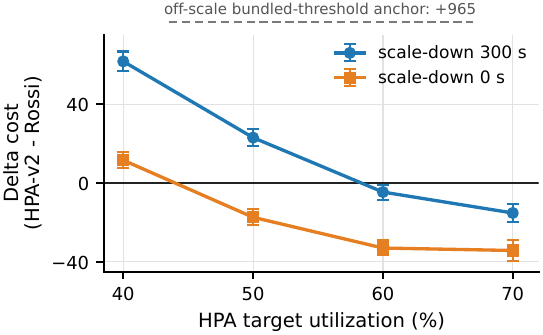}
\caption{Rossi's lag-cell advantage across HPA-v2 configurations. Every configuration sits far below the bundled-threshold collapse (dashed, $+965$). Rossi retains the cell only at the most cautious settings, a $40\%$ target or $50\%$ with full stabilisation. At a $50\%$ target without stabilisation, and at any target of $60\%$ and above, the production controller wins even under lag. The grid maximum of $+62$ puts the bundled comparison's overstatement above $15\times$ at every configuration.}
\label{fig:hpa-config}
\end{figure}

\paragraph{Robustness to the controller's configuration}
The $50\%$ target and $300\,$s window of \Cref{tab:hpa} are one configuration, and the outcome could depend on it. We therefore varied the target utilisation over $\{40,50,60,70\}\%$ and the scale-down window over $\{300,0\}\,$s and re-ran all four cells. Rossi retains the lag cell only at the most cautious settings, $+62$ at a $40\%$ target with full stabilisation, $+23$ at the reported point, and $+12$ at $40\%$ without stabilisation. At a $50\%$ target without stabilisation, and at every target of $60\%$ or above, the production controller wins even under lag (\Cref{fig:hpa-config}). Across all eight configurations, the lag-cell divergence never exceeds $+62$ cost units, at least fifteen times below the bundled threshold's $+965$ collapse. The only setting in which Rossi regains the clean cell is the most cautious one, a $40\%$ target with full stabilisation, where the adversarial cell also becomes the grid's only zero-straddling interval. 
The artefact is the bundled comparator's collapse, not the production controller's tuning, and no configuration reproduces it.

\paragraph{The same check reframes the Decima result}
The bundled Decima comparator, \texttt{dynamic\_partition}, has the symmetric weakness of never reallocating. We re-ran the four Decima cells against a dependency-aware shortest-remaining-work (SRW) scheduler that selects the arrived job with the least estimated remaining work, then its least-remaining ready node, and allocates executors work-conservatively. Decima beats this stronger comparator in all four cells, including under lag, by $1.4\times10^{5}$ (clean), $1.8\times10^{6}$ (lag), $4.1\times10^{5}$ (tail), and $1.5\times10^{5}$ (adversarial) mean-JCT units. These results delimit what the Decima cells are evidence of. They are not evidence that a better baseline overturns Decima, because against two classical schedulers it is not overturned. They are evidence of the other failure documented here. The comparison the published headline rests on, against tuned heuristics on a live testbed, is absent from the released artefact, so the reported $21\%$ advantage cannot be evaluated from it. Against the strongest classical comparator its reference evaluation exposes, the margin is $3.0\%$. A method can be strong while its central published claim remains unverifiable. Decima exhibits both, and only the second is a claim this paper needs.

\subsection{Magnitude sensitivity}
\label{sec:exp:magnitude}

\begin{figure}[!t]
\centering
\includegraphics[width=\columnwidth]{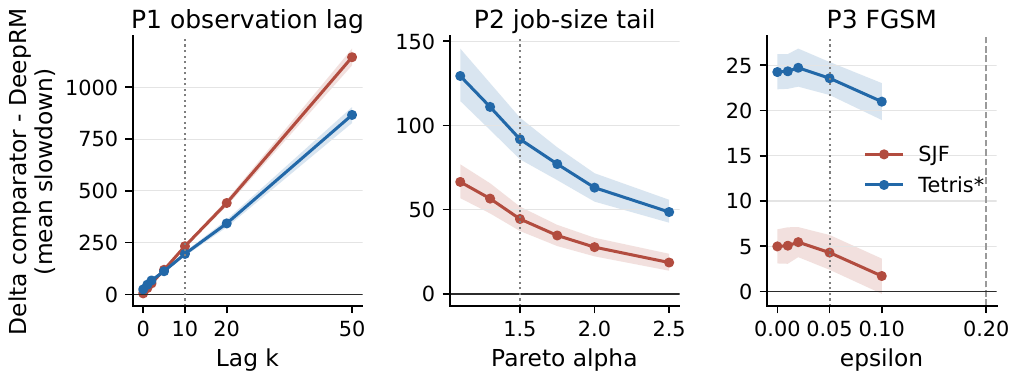}
\caption{DeepRM under the pre-registered magnitude sweeps, against the locked Tetris$^\star$ and SJF comparators. The falsification direction is stable across the realistic range, the learned policy is ahead at every lag and every adversarial magnitude, and the divergence widens as the comparators degrade. The severe point $\varepsilon=0.2$ is a non-terminating evaluation and is omitted. $\Delta=\text{metric}(\text{comparator})-\text{metric}(\text{DeepRM})$, and the dotted line marks the calibrated anchor.}
\label{fig:deeprm-sweep}
\end{figure}

\begin{figure}[!t]
\centering
\includegraphics[width=\columnwidth]{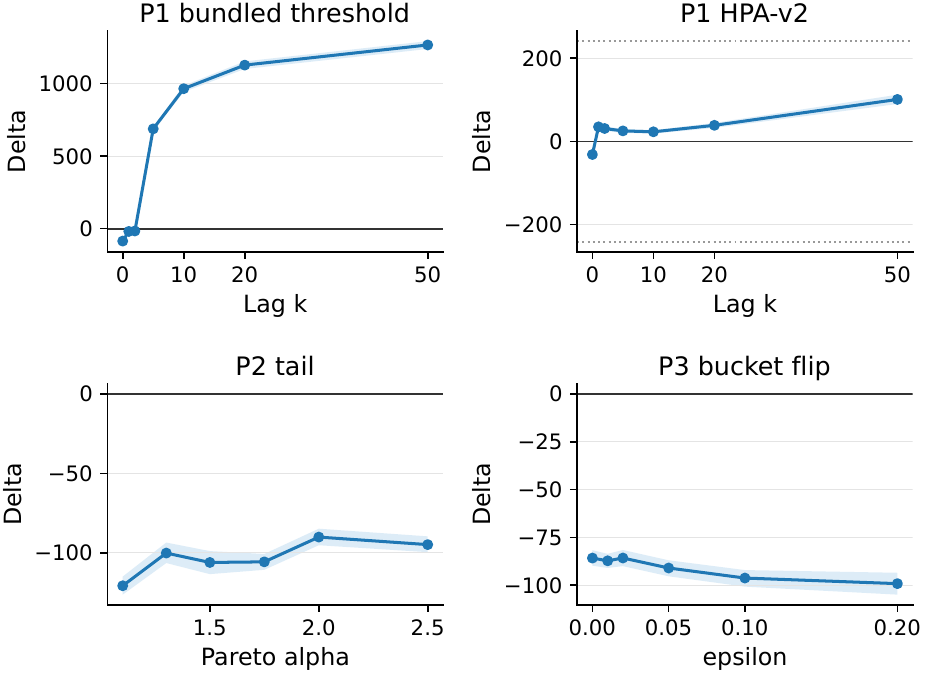}
\caption{Rossi across the pre-registered perturbation sweeps. The bundled threshold diverges for $k\ge5$, while HPA-v2 remains close to Rossi across the lag grid (top). The tail and adversarial verdicts remain stable in sign (bottom). Dotted lines mark the $25\%$ alarm threshold. $\Delta=\text{metric}(\text{comparator})-\text{metric}(\text{Rossi})$.}
\label{fig:rossi-sweep}
\end{figure}

A conclusion drawn at a single anchor per perturbation could depend on the choice of that anchor. We therefore pre-registered a sweep of each perturbation over a grid of magnitudes and re-evaluated the fixed policies at every point. We committed in advance to report any verdict that reverses inside the realistic range, even where the reversal favours the explanation we argue against. Across the three methods the anchor verdicts are stable in sign with one exception. DeepRM stays ahead of both bundled comparators at every lag and adversarial magnitude in the realistic range, with the difference widening as the comparators degrade (\Cref{fig:deeprm-sweep}), and the one severe point, $\varepsilon=0.2$, drove a non-terminating rollout and is reported as noncompletion outside the confirmatory range. For Rossi the result of \Cref{sec:exp:hpa} strengthens. The HPA-equivalent controller stays within a small band of the learned policy across the entire lag grid. Its largest difference, about $101$ cost units, is a tenth of the bundled threshold's $k=10$ collapse and stays below the pre-registered $25\%$ alarm. The bundled collapse is itself an onset, with the threshold competitive at $k\le2$ and diverging only from $k\ge5$ (\Cref{fig:rossi-sweep}). The Rossi tail and adversarial verdicts are sign-stable across their grids.

\begin{figure}[!t]
\centering
\includegraphics[width=\columnwidth]{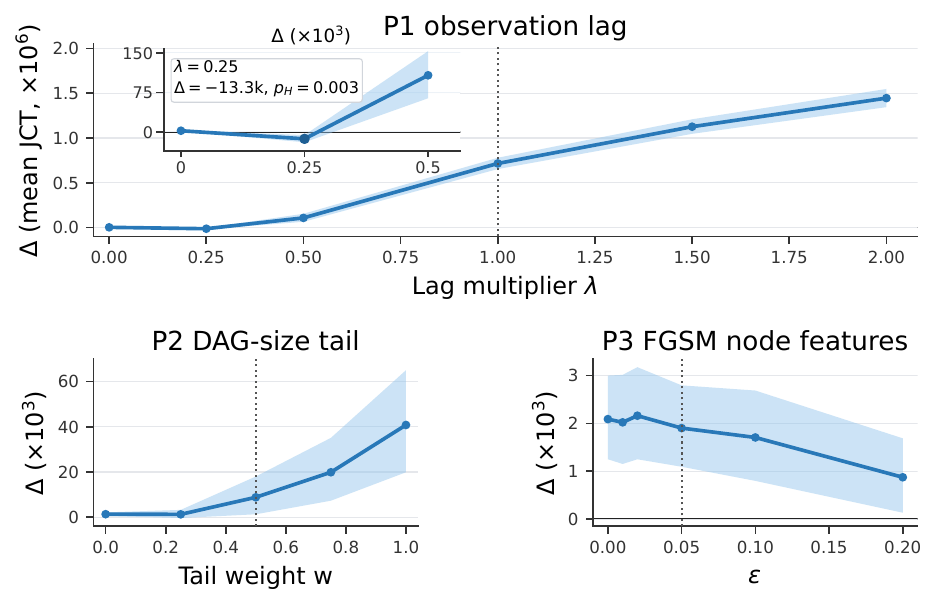}
\caption{Decima under the pre-registered magnitude sweeps, against the locked \texttt{dynamic\_partition} comparator. P2 and P3 are sign-stable, with the P2 margin growing past its marginal anchor and the P3 falsification holding across the full $\varepsilon$ grid. P1 is not. The inset shows the pre-registered $\lambda=0.25$ cell, where the comparator is significantly cheaper ($\Delta=-13{,}305$, CI $[-19{,}833,-6{,}168]$, $p_{\mathrm{H}}=0.003$, 24 of 30 seeds) while every other magnitude favours Decima by up to $1.4\times10^{6}$ JCT units. A verdict that reverses on the choice of a single magnitude is the failure mode single-anchor evaluation cannot detect. $\Delta=\text{JCT}(\texttt{dynamic\_partition})-\text{JCT}(\text{Decima})$, and dotted lines mark the calibrated anchors.}
\label{fig:decima-sweep}
\end{figure}

The exception is Decima under observation lag, reported as registered (\Cref{fig:decima-sweep}). The verdict is not sign-stable inside the calibrated range. At $\lambda=0.25$ the static comparator is significantly cheaper than Decima, with a Holm-adjusted $p$ of $0.003$ and 24 of 30 paired seeds favouring the comparator, while at the $\lambda=1.0$ anchor and beyond Decima leads by large margins. We audited the lag sampler and found no rounding artefact, so the reversal is not a sampler artefact. Because each magnitude draws its own perturbation rather than a scaled version of one draw, it is not a point on a smooth response curve, and we make no monotonicity claim for this cell. We therefore no longer state the Decima lag result as a magnitude-robust falsification, and the claim we retain is its anchor-level robustness against stronger comparators (\Cref{sec:exp:hpa}). Decima's other grids carry no such reversal. P3 is sign-stable across the full $\varepsilon$ range, and P2 is marginal at the anchor and Holm-significant from $w\ge0.75$. A pre-registered comparison that changes sign on the choice of a single magnitude demonstrates that single-anchor evaluation cannot settle the question it asks.

\subsection{Two supporting analyses}
\label{sec:exp:ablation}

\emph{Autocorrelation sensitivity (Rossi).} The 30 Rossi windows are slices of one trajectory, so the paired bootstrap's independence assumption is only approximate. A pre-registered block-bootstrap analysis confirms the verdicts. Lag-1 autocorrelations are below $0.25$ in all four cells, no Ljung--Box independence null rejects, and moving-block intervals at $L\in\{5,10\}$ never change a cell's zero-containment status. The block sign-flip test is uninformative at $N=30$, because its discreteness floors the $p$-value at $0.125$ for $L=10$, so the moving-block intervals are the better summary. In the P1 cell those intervals are centred more than $100$ standard errors from zero.

\emph{Action-redundancy ablation (DeepRM).} The decision-versus-outcome separation in P3 invites a redundancy explanation, in which many near-equivalent actions per state mean that flipping the argmax rarely selects a catastrophic alternative. We test this by training DeepRM under restricted visible action sets $M\in\{10,3,1\}$ at the source-aligned budget. Reducing $M$ from $10$ to $3$ keeps DeepRM competent against the matched comparator and produces no detectable increase in FGSM aggregate degradation, with $\text{deg}_3$ and $\text{deg}_{10}$ both statistically indistinguishable from zero and the observed ratio opposite in sign to the prediction. The $M=10$ condition reproduces the locked P3 aggregate exactly while sampling a different diagnostic state set, which is why its argmax-change rate ($75.2\%$) differs from the locked diagnostic ($69.8\%$, \Cref{sec:exp:pattern}). The $M=1$ condition fails the competency gate, which reflects training failure under extreme action restriction rather than a robustness mechanism. The redundancy explanation is not supported. The structural origin of the separation remains open, and that is exactly the kind of question a typed action space (\Cref{sec:bg:actions}) would make tractable by construction.

\subsection{Limitations}
\label{sec:exp:limits}

The findings hold at the anchors tested, and \Cref{sec:exp:magnitude} shows the verdict signs are stable across magnitude in every cell but one, the Decima lag cell, whose sign we no longer claim as magnitude-robust. The scope is three methods, and two of them, DeepRM and Decima, share an author lineage. Each was selected as the canonical, most built-on instance of its task class, and breadth across groups is future work. The structural origin of the action-versus-outcome robustness difference (\Cref{sec:exp:pattern}) is unresolved. Most consequentially, the public simulators capture queueing, packing, and arrivals but not the contention, tail-latency dynamics, micro-bursts, or scheduling jitter that dominate production performance. This bounds what the programme can claim, and it does so asymmetrically. It would undercut an absolute-performance claim, which we do not make. It does not undercut the claims we do make, because both kinds are independent of simulator fidelity. Whether an artefact can express the comparison its headline rests on is a property of the artefact, not of its fidelity, and a relative comparison between two policies on the same trajectory holds whatever that trajectory omits. Within this bound, three results remain. The pre-registered predictions do not survive against the bundled comparators, the strongest result attenuates by more than an order of magnitude against a production-grade controller robust to its configuration, and a second method withstands stronger comparators at its anchor while its published headline remains unverifiable.

\section{Discussion}
\label{sec:discussion}

\subsection{Interpreting the adversarial correspondence}
Production telemetry carries the magnitude of bounded observation perturbations, workload shifts have the structure of environment poisoning, and scalar proxies satisfy the preconditions of reward poisoning, so the correspondence in \Cref{tab:adversarial} describes operating conditions rather than hypothetical attacks. Our use of the adversarial-RL literature~\cite{huang2017adversarial,gleave2020adversarial,zhang2020robust,rakhsha2020policy,behzadan2017vulnerability} is a reframing rather than a new attack or defence, and \Cref{sec:evidence} subjects that reframing to a pre-registered test on canonical orchestration methods. The test refutes the directional claim. The learned policy is not the party that degrades against the comparators the field uses, and the strongest case for it dissolves against a production-grade baseline. 

\subsection{Key performance indicators selection}
The Rossi cell is a small instance of a general failure, in which a weighted-sum reward becomes the metric and the weights, not the behaviour, set its scale. The published cost weights SLA compliance at $0.90$, resource cost at $0.09$, and reconfiguration churn at $0.01$, pricing churn two orders of magnitude below the SLA term. The released traces show the consequence. In the clean cell the learned policy issues roughly $1{,}500$ reconfigurations per window against the threshold controller's $56$, a $27\times$ difference the scalar renders invisible. Re-pricing the churn to a still-modest $0.30$ widens the threshold controller's clean-cell advantage over the learned policy from $99$ to $487$ cost units, a difference a reader of the published number never sees. Across the full re-pricing grid no cell's verdict flips, so the weight chooses the margin rather than the winner. A scalar cost hides the trade-off it encodes. Orchestration results should be reported in operational terms, with SLA violations, action churn, overload peaks, and tail latency kept distinct, so that a reader can apply the weighting their own deployment implies rather than inherit the authors'.

\subsection{Limits of benchmark evidence}
Current orchestration benchmarks model queueing, packing, and arrival processes but omit resource contention, cross-tenant interference, microbursts, scheduling jitter, and component failures. Performance gains measured without these conditions do not establish deployment readiness. A credible benchmark should include failure scenarios, delayed and noisy telemetry, heavy-tailed and bursty workloads, tenant churn, and bounded observation perturbations. It should also compare learned controllers with production-grade baselines.
%
\subsection{Scope and requirements for valid evaluation}
This paper neither proposes a comprehensive evaluation standard for learned orchestration nor presents a method that satisfies such a standard. It puts an effort into isolating the deficiencies that prevent current studies from supporting deployment-relevant conclusions and identifies the evidence that a valid evaluation would require.
We argue that a valid evaluation requires four technical components. 
First, the evaluation infrastructure should combine a multi-tenant testbed with extensible simulation or emulation. End-to-end application suites such as DeathStarBench provide realistic service dependencies and request paths~\cite{gan2019open}, while trace-driven frameworks such as faas-sim support controlled evaluation of scaling, placement, routing, and scheduling across heterogeneous containerised and serverless infrastructures~\cite{raith2023faassim}. Physical execution is still required to expose contention, scheduling jitter, and other effects that cannot be validated exclusively through simulation.
Second, the workload corpus must cover several distributional and temporal profiles derived from production traces~\cite{reiss2012heterogeneity,cortez2017resource}. Third, the comparator set must include both academic methods and deployed controllers, including DeepRM, Decima, default Kubernetes scalers, Autopilot's recommenders~\cite{rzadca2020autopilot}, and structurally constrained methods such as HPAQT~\cite{mayerhofer2025hpaqt}. Fourth, results must be reported by task class and by operational metric, in alignment with \Cref{tab:taxonomy}.

%
These components also require changes to publication and review criteria. Evaluations should justify comparator selection and perturbation models before testing, report operational metrics separately, release sufficient material for reproduction, and treat negative results as evidence. Absent these criteria, successive benchmark improvements remain isolated results rather than cumulative evidence of operational validity.
\subsection{Outlook}
\label{sec:discussion:outlook}
The findings above do not imply that learning has no role in orchestration. They indicate that the formulation should begin with the system's operational structure rather than a benchmark-defined Markov decision process. One proposed direction is a computing-continuum system defined by application requirements, observable metrics, infrastructure configurations, and admissible adaptations~\cite{dustdar2023continuum}. A Markov blanket links these elements, while equilibrium denotes an operating state consistent with the system requirements and available infrastructure. Under this formulation, learning estimates system dependencies and selects reconfigurations within a structure defined before training. It does not determine the operational objective or invent the actions through which the system is controlled.

Danilenka \emph{et al.} provide an initial instantiation of this direction in heterogeneous and lifelong federated learning~\cite{danilenka2024adaptive}. Their agents represent SLO fulfilment, configuration variables, and system measurements in a learned Bayesian network, then select configurations under data drift and hardware heterogeneity. The physical-testbed results show that high-level SLOs and explicit configuration choices can be combined with a learned environment model. They do not establish that active inference solves the service orchestration problem. The application scope is narrower, the configuration space is enumerated, and scalability, temporal modelling, and more complex SLO specifications remain open.
Extending this direction to orchestration requires three commitments. Objectives should follow from operational contracts, represented as hard constraints through constrained MDPs~\cite{altman1999} or as preferred outcome distributions, rather than as freely weighted scalar rewards. Actions should be typed against control-plane interfaces, including their preconditions, costs, reversibility, and feasibility. Evaluation should compare learned controllers with production-grade controllers under registered perturbations, with SLO violations, reconfiguration churn, overload, and tail latency reported separately. Robust-MDP methods~\cite{iyengar2005robust,nilim2005robust,wiesemann2013robust,pinto2017robust} provide one way to formalise uncertainty, but they do not determine the correct objective, action space, or evidence standard. 
Therefore, progress requires methods whose objectives and admissible actions are grounded in system contracts and whose operational validity is established against production-grade controllers under registered perturbations.
\section{Conclusion}
\label{sec:conclusion}
We examined whether production conditions explain why reinforcement learning has not reached service orchestration at scale. Across three influential systems, the registered tests did not produce a stable verdict on learned control. Several apparent advantages arose from weak comparators or perturbation implementations. One advantage decreased by roughly fortyfold against a production-grade controller. Another published claim could not be reconstructed from the released artefact. Other conclusions changed with perturbation magnitude or evaluation mode. These results do not establish that learned controllers withstand production conditions, but demonstrate that the available comparisons cannot reliably determine whether they do.

Based on these results and the broader literature, we identify a methodological and institutional problem. Research and review incentives favour benchmark improvements over available comparators, even when the comparison lacks an operational win condition and provides little evidence of deployment performance. This prevents results from accumulating into a reliable account of when learning improves orchestration. Addressing this problem requires changes to both evaluation practice and how contributions are judged. Learned controllers should be compared with production-grade controllers under registered perturbations. Operational outcomes should be reported separately rather than combined through author-selected scalar weights. Released artefacts should reproduce the claims for which a method is cited. Objectives and actions should also follow from system contracts and control-plane interfaces rather than from benchmark conventions. These requirements do not determine which learning method should replace current controllers. They establish the evidence needed to decide whether any method should be used. Until publication incentives reward production-relevant evidence, additional funding, computation, and publication volume will expand the literature without establishing whether reinforcement-learning-based anticipatory control improves service orchestration under production conditions.

\section*{Acknowledgment}
We thank Alexander Knoll for providing us with the necessary hardware infrastructure. This work has been supported by the Marie Skłodowska-Curie Actions Doctoral Networks (DN) under call HORIZON-MSCA-DN-2024-01-01 (Project ``SAILING'') .

\balance
\bibliography{references}

\end{document}